\documentclass[aps,prl,reprint,nofootinbib,superscriptaddress]{revtex4-2} 
\usepackage{graphicx,url,amssymb,amsmath,rotating,color,units,wasysym,epsfig,multirow,epstopdf, stackengine}
\usepackage{amsmath}
\usepackage{cancel}
\usepackage{physics}
\usepackage{appendix}
\usepackage{amssymb}
\usepackage[colorlinks,urlcolor=blue,citecolor=blue,linkcolor=blue]{hyperref}
\usepackage[normalem]{ulem}
\usepackage{tabularx}
\usepackage{tensor}
\newcolumntype{Y}{>{\centering\arraybackslash}X}

\usepackage{soul}

\usepackage[commentmarkup=uwave]{changes}

\definechangesauthor[name=EP, color=cyan]{EP}
\definechangesauthor[name=MI, color=teal]{MI}
\definechangesauthor[name=WF, color=blue]{WF}
\definechangesauthor[name=YC, color=orange]{YC}

\newcommand{\change}[1]{{#1}}

\newcommand{\CIT}{\affiliation{Department of Physics, California Institute of Technology, Pasadena, California 91125, USA}}
\newcommand{\CITLab}{\affiliation{LIGO Laboratory, California Institute of Technology, Pasadena, California 91125, USA}}
\newcommand{\CCA}{\affiliation{Center for Computational Astrophysics, Flatiron Institute, 162 5th Ave, New York, New York 10010, USA}}
\newcommand{\StonyBrook}{\affiliation{Department of Physics and Astronomy, Stony Brook University, Stony Brook NY 11794, United States}}
\newcommand{\PI}{\affiliation{Perimeter Institute for Theoretical Physics, 31 Caroline St., Waterloo, ON, N2L 2Y5, Canada}}

\begin{document}

\title{The curvature dependence of gravitational-wave tests of General Relativity}

\author{Ethan Payne}
\email{epayne@caltech.edu}
\CIT
\CITLab

\author{Maximiliano Isi}
\email{misi@flatironinstitute.org}
\CCA

\author{Katerina Chatziioannou}
\email{kchatziioannou@caltech.edu}
\CIT
\CITLab

\author{Luis Lehner}
\email{llehner@perimeterinstitute.ca}
\PI

\author{Yanbei Chen}
\email{yanbei@caltech.edu}
\CIT

\author{Will M. Farr}
\email{wfarr@flatironinstitute.org}
\CCA
\StonyBrook

\begin{abstract}
High-energy extensions to General Relativity modify the Einstein-Hilbert action with higher-order curvature corrections and theory-specific coupling constants.
The order of these corrections imprints a universal curvature dependence on observations while the coupling constant controls the deviation strength.
In this Letter, we leverage the theory-independent expectation that modifications to the action of a given order in spacetime curvature (Riemann tensor and contractions) lead to observational deviations that scale with the system length-scale to a corresponding power.
For gravitational-wave observations, the relevant scale is the binary total mass, and deviations scale as a power of mass $p$ related to the action order.
For example, $p=4,6$ arise in effective field theory for cubic and quartic theories respectively.
We incorporate this \change{insight} into the inspiral phase test of General Relativity with current gravitational-wave observations, \change{and directly infer} the curvature scaling without compromising the agnostic nature of the test.
This introduces a flexible yet highly interpretable new paradigm for \change{tests of General Relativity spanning many length scales}.
\end{abstract}

\maketitle

\section{Introduction}

Searches for deviations from General Relativity (GR) with gravitational waves (GWs) are hampered by the vast landscape of alternative theories~\cite{Yunes:2013dva,Shankaranarayanan:2022wbx} and the scarcity of detailed predictions under any specific theory.
Faced with these challenges, most tests are framed as theory-agnostic searches for generic deviations~\cite{LIGOScientific:2016lio,LIGOScientific:2018dkp,LIGOScientific:2019fpa,LIGOScientific:2020tif,LIGOScientific:2021sio}.
Although this approach has provided increasingly precise null tests, it forgoes physical expectations for the likely behavior of realistic deviations, making constraints harder to interpret and potentially less sensitive~\cite{Perkins:2022fhr,Mehta:2022pcn}.
However, even without reference to a specific beyond-GR theory, general arguments limit how deviations may manifest under broad theory classes.
This presents an opportunity for improving tests of GR with GWs.

In this paper, we exploit one such argument arising from effective field theory: the magnitude of the beyond-GR effect should scale with the spacetime curvature of the source, e.g.,~\cite{Donoghue:2017ovt,Burgess:2020tbq}.
Since curvature is proxied by mass, \emph{lighter} systems should manifest \emph{larger}---and hence more measurable---deviations.
In the context of binaries observed with GWs, this expectation has been folded in (though not directly inferred from) post-merger (ringdown) constraints~\cite{Carullo_2021, Maselli:2019mjd, Maselli:2023khq} 
and simulations of residual cross-correlated power between detectors~\cite{Dideron:2022tap}.
Here, we \change{go beyond folding in fixed values of the curvature scaling~\cite{Carullo_2021, Maselli:2019mjd, Maselli:2023khq}: we} exploit the fact that the curvature scaling will manifest within an ensemble of observations to \emph{directly infer} the curvature dependence of GR extensions without resorting to theory-specific assumptions.
\change{A direct measurement of the curvature scaling will then provide insights on the modification to the Einstein-Hilbert action.}

We propose a search for deviations from GR in a catalog of GW observations that leverages this effective field theory insight.
Instead of committing to a specific theory, we constrain expected morphologies from a large set of potential theories at once.
Deviations from GR are linked to the leading power correction in the Einstein-Hilbert action, e.g., \cite{Weinberg_2008,Stein:2013wza,Endlich:2017tqa,Bueno_2017,Cano:2021myl,Cano:2022wwo}, with a Lagrangian
\begin{equation} \label{eq:lagrangian}
{\cal L} = {\cal L}_\textrm{GR} + \lambda \,F_\gamma({\cal{R}},\phi)\,,
\end{equation}
where ${\cal L}_\textrm{GR}$ is the GR term, $F_\gamma({\cal{R}},\phi)$ is some functional of the curvature ${\cal{R}}$ (and potentially other
degrees of freedom $\phi$) scaling as ${\cal{R}}^\gamma$, and $\lambda$ is the dimensionful coupling coefficient.\footnote{The functional $F_\gamma({\cal{R}}, \phi)$ could include any combination of curvature tensors, and/or their derivatives (not just the Ricci scalar $R$), derivatives of $\phi$, and couplings that scale as $\ell^{-2\gamma}$~\cite{Stein:2013wza}.}
Dimensional analysis reveals that $\lambda\sim\ell^{2(\gamma-1)}$, representing a theory-specific coupling governed by a theory-specific length-scale, $\ell$.
Importantly, this scaling is imprinted in (dimensionless) deviations
from GR {\em regardless of the physical mechanism} they induce.

For instance, consider beyond-GR theories with cubic or quartic curvature corrections, $\gamma=\{3,4\}$, and no further degrees of freedom.
Such theories introduce tidal effects in black hole binaries, under the assumption that the theory-specific length scale is smaller than the lightest black hole.
Such deviations first appear at the 5th post-Newtonian (PN) order through tidal
Love numbers whose magnitude depends on the specific correction.\footnote{Nominally 2PN effects are also introduced, but their
contribution is subleading to tidal effects if $\ell \lesssim 5$\,km.}
Crucially, these deviations scale as $\lambda/M^{2(\gamma-1)} \propto M^{-\{4,6\}}$~\cite{Cardoso2018,Cano:2021myl,Cayuso:2023xbc}, with $M$ the binary total mass. 
Additional degrees of freedom impact these scalings. 
For instance, quadratic theories, 
$\gamma=2$, with 
additional degrees of freedom yield 
hairy black holes with deviations in the inspiraling GW signal at either $-1$ or 2PN order depending
on the parity of the correction~\cite{Yagi:2011xp}.
Now the dimensionless deviation is  $(\lambda/M^{2(\gamma-1)})^2 \propto M^{-4}$ with the additional square power coming from the coupling of the scalar degree of freedom and the metric tensor.
In either case, constraining the value of the mass exponent has tremendous power in narrowing viable corrections.

In this Letter, we exploit the expected curvature/mass scaling in the context of deviations in the post-Newtonian inspiral phase of binary black-hole coalescences.
We incorporate the mass dependence into hierarchical tests of GR and infer both the magnitude and the curvature dependence of measured deviations.
Using observations from the third LIGO-Virgo-KAGRA~\cite{LIGO,Virgo,KAGRA} GW transient catalog~\cite{gwtc2, gwtc3}, we confirm the validity of GR.
We further highlight the method's ability to constrain the curvature order at which a modification appears with simulated observations.

\section{Constraining the curvature dependence with gravitational waves}

Combining information from a catalog of GW observations in a theory-agnostic way amounts to characterizing the distribution of putative deviations~\cite{Zimmerman:2019wzo,Isi:2019asy,Isi:2022cii}, with GR recovered under vanishing deviations for all sources.
This hierarchical framework can incorporate arbitrary numbers of deviation parameters~\cite{Zhong:2024pwb}, astrophysical parameters~\cite{Payne:2023kwj}, and selection effects~\cite{Magee:2023muf}.
In all cases so far, the deviation population has been modeled as a (potentially multidimensional) Gaussian, whose mean $\mu$ and variance $\sigma^2$ are global parameters, independent of source properties.
This framework provides a powerful null test of GR, which is recovered for $\mu = \sigma =0$, see Ref.~\cite{Pacilio:2023uef} for inference caveats; however, it does not impose any structure on the scale of the deviations as a function of source parameters.

We extend the hierarchical framework to incorporate the expectation that the magnitude of deviations scales with source curvature by anchoring the deviation distribution to the total binary (source-frame) mass $M$.
We achieve this by reparametrizing $\mu$ and $\sigma$ as
\begin{equation}
    \mu = \mu_0\left(\frac{M}{10\,M_\odot}\right)^{-p}\,,\quad \sigma = \sigma_0\left(\frac{M}{10\,M_\odot}\right)^{-p}\,,\label{eq:musigma}
\end{equation}
where $\mu_0$ and $\sigma_0$ control the magnitude of the conditional mean and spread of the GR deviation at $M=10\,M_\odot$.
The curvature scaling order, $p$, is directly related to the index, $\gamma$ in Eq.~\eqref{eq:lagrangian}, as either $p=2(\gamma-1)$ in the absence of additional fields or $p=4(\gamma-1)$ in their presence.
In this notation, $p=4$ corresponds to quadratic curvature corrections with additional degrees of freedom or cubic corrections in their absence while $p=6$ implies quartic corrections.
Propagation effects, such as modifications to the GW dispersion relation~\cite{Will:1997bb, Mirshekari:2011yq}, impose source-independent deviations and, thus, $p=0$.
Even when the deviation distribution is not Gaussian, this method will still identify a violation of GR~\cite{Isi:2022cii} and, if the shape of the distribution is unchanged over the binary mass range, also identify the scaling, $p$.
Irrespective of $p$, GR corresponds to $\mu_0 =\sigma_0 = 0$.

While this framework can be applied to any test that infers both the deviation and the system total mass (such as correlated power among detectors~\cite{Dideron:2022tap,Dideron_inprep}), we turn our attention to the post-Newtonian phase deviation test~\cite{Yunes:2009ke,Sampson:2013lpa,Li2012PN, agathos2014TIGER, Mehta:2022pcn}.
Deviations at the $(k/2)$PN order are inferred by varying the respective phasing coefficient by some (dimensionless) fractional deviation, $\delta\varphi_k$~\cite{agathos2014TIGER}.
Since GR is recovered when $\delta\varphi_k=0$ and the parameter is dimensionless, this is a null test that should follow a curvature scaling as we have described.
Below, we consider deviations from $-1$PN up to $3.5$PN order, including logarithmic terms~\cite{gwtc2, gwtc3}.

Following Refs.~\cite{gwtc2, gwtc3, LIGOScientific:2021sio, Payne:2023kwj}, we consider the 20 observations from the third LIGO-Virgo observing run with a false-alarm-rate of less than $1/1000\,\textrm{yr}$ and with
an estimated inspiral signal-to-noise ratio greater than 6 \change{(Tab. I of Ref.~\cite{Payne:2023kwj})}. 
\change{We do not consider data from the first and second observing periods as semi-analytic simulations for sensitivity estimation are not available.} 
Individual-event posteriors
were computed in Refs.~\cite{gwtc2, gwtc3, LIGOScientific:2021sio} with a modified form of the {\tt \sc SEOBNRv4}
waveform~\cite{bohe2017, Cotesta2018, Brito2018, Cotesta2020, Mehta:2022pcn} and released in Refs.~\cite{gwtc2_tgr_release, gwtc3_tgr_release}. To
mitigate against systematic bias due to incorrect astrophysical assumptions, we
jointly model the distribution of the GR deviation parameter and the system
masses and spins with the astrophysical population
models and selection function~\cite{Payne:2023kwj}. Based on Ref.~\cite{Magee:2023muf}, we assume
that there are no direct selection effects for the magnitude of the deviation. We infer the population distribution of each
post-Newtonian term separately with uniform hyperpriors on $\mu_0\in[-30,30]$, $\sigma_0\in[0,100]$, and $p\in [-1,8]$, chosen so as to remain agnostic on the magnitude and character of the curvature scaling.

The expected inference structure depends on a number of considerations.
Observations of signals spanning $10{-}100\,M_\odot$ in total mass~\cite{gwtc2, gwtc3} have yielded no evidence for a violation of GR~\cite{LIGOScientific:2021sio}. 
Among those, constraints are generally stronger for lighter signals with more inspiral cycles~\cite{Mishra:2010tp}, however, there are more observed signals at $M\sim60\,M_{\odot}$~\cite{gwtc3_pop}.
Importantly, for \change{$p\geq4$} lighter systems are expected to manifest larger deviations. 
Absent a detected deviation, we expect those systems to provide the overall strongest constraints.
This expectation plays out in the results below.

\begin{figure}
    \centering
    \includegraphics[width=\linewidth]{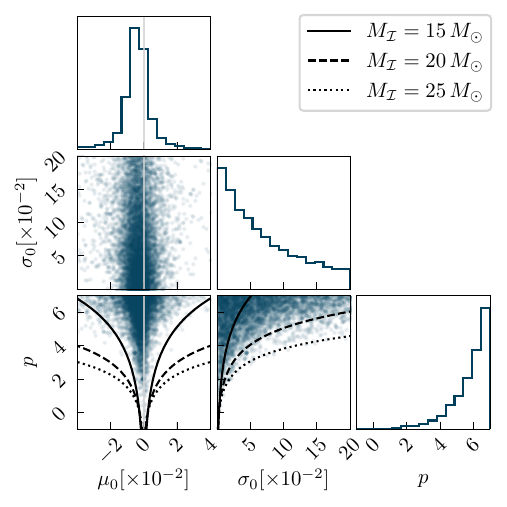}
    \caption{Posterior distribution for the $-1$PN deviation population parameters inferred from the 20 GW observations in GWTC-2 and GWTC-3 which pass the threshold criteria~\cite{LIGOScientific:2021sio, gwtc2, gwtc3, Payne:2023kwj},
    confirming consistency with GR, $(\mu_0,\sigma_0)=(0,0)$.
    Due to the non-detection of a violation, the constraint is dominated by $M_{\cal I} \in [15,25]\,M_\odot$ and the posterior is bounded per Eq.~\eqref{eq:MI} (lines).
    While the marginal posterior for the scaling parameter, $p$, indicates preference for larger values, it is a product of this bounded structure.}
    \label{fig:gwtc3_corner}
\end{figure}

Figure~\ref{fig:gwtc3_corner} shows results for the $-1$PN deviation, related to deviations due to a scalar field coupling to the Gauss-Bonnet invariant, i.e., Einstein-scalar Gauss-Bonnet~\cite{Yagi:2011xp}. 
We discuss this order in detail, but obtain qualitatively similar results for other PN orders. 
The constraints are consistent with $(\mu_0,\sigma_0)=(0,0)$ and, thus, GR.

To further understand the posterior, consider that, in the absence of inferred deviations and denoting the most informative mass range as $M_{\cal I}$, the allowed values of $\{\mu_0, \sigma_0, p\}$ correspond to deviations that would be undetectable at $M_{\cal I}$:
\begin{equation}
    \{\mu_0, \sigma_0\} \left(\frac{M_{\cal I}}{10\,M_\odot}\right)^{-p} \sim \textrm{const.}\,,\label{eq:MI}
\end{equation}
where the constant represents the test sensitivity.
To determine $M_{\cal I}$, we split events based on their total mass into 5$\,M_{\odot}$ bins and compute the \emph{precision}
\begin{equation} \label{eq:prec}
    {\cal{P}}(M,p)\equiv\frac{1}{\Sigma^2(M,p)} = \sum_{i=1}^{N_\textrm{b}}\frac{1}{\Sigma_i^2\Big(\frac{M_i}{10\,M_\odot}\Big)^{2p}}\, ,
\end{equation}
where $i$ runs over the $N_\textrm{b}$ events within the bin with central mass $M$, $\Sigma_i^2$ is
the variance of the GR deviation of an individual event marginalized over all other parameters, and $M_i$ is the median total mass. 
The precision corresponds to the total inverse variance scaled by the expected value of the deviation in each mass bin; it therefore quantifies which mass range is more constraining. 
For the $-1$PN order, the precision is maximized when $M_{\cal I} \in [15,25]\,M_\odot$, resulting in the black lines in Fig.~\ref{fig:gwtc3_corner}, which track the general shape of the posterior.

\begin{figure*}
    \centering
    \includegraphics[width=\linewidth]{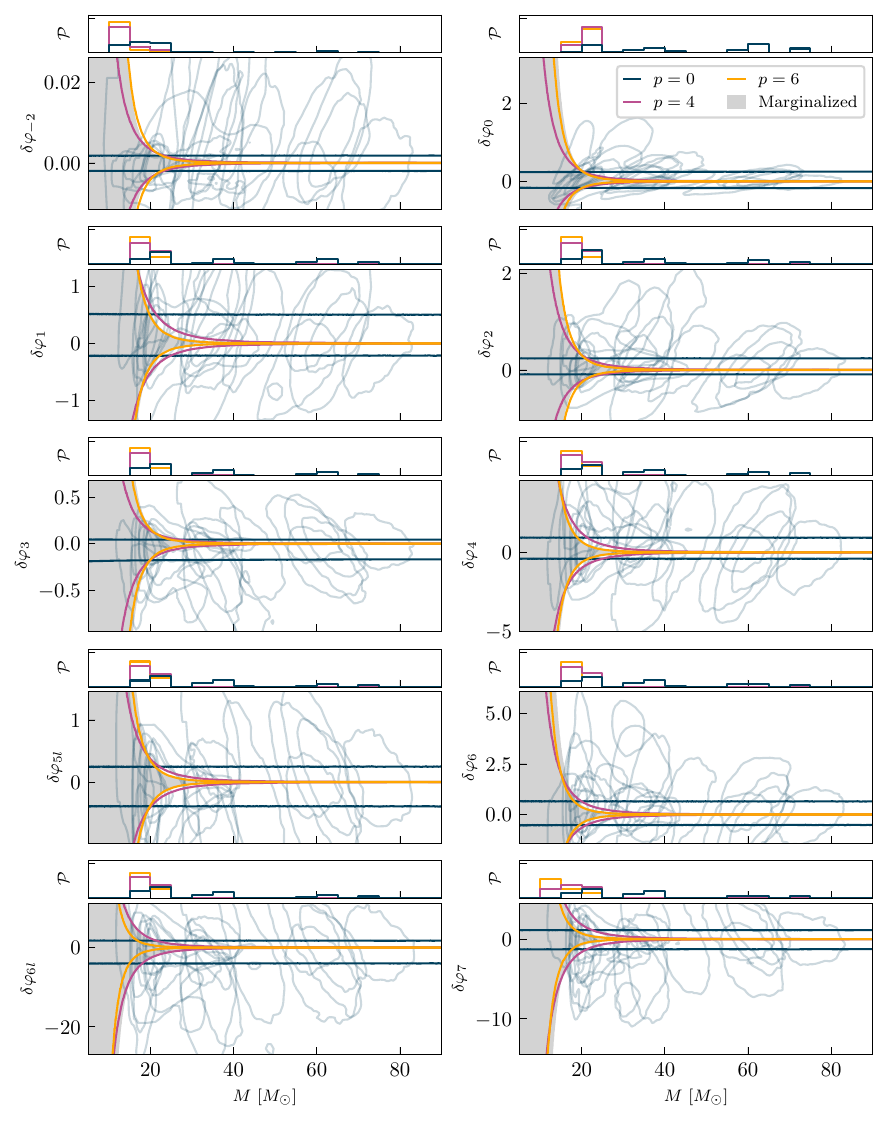}
    \caption{Posterior predictive distributions, Eq.~\eqref{eq:PPD}, for deviations across all PN orders (main panel) and the precision, ${\cal{P}}$ (top sub-panel), as a function of binary total mass. We show results for fixed values of $p=0,4,6$ indicative of different theoretical models and when marginalizing over $p$. The \change{90\% credible regions of the} 20 individual-event posteriors are shown in faint blue.
    \change{The precision indicates the relative contribution on the constraints for the different curvature orders, generally maximized at $\sim20\,M_\odot$; it is normalized for each $p$}. 
    }
    \label{fig:ppds}
\end{figure*}

Equation~\eqref{eq:prec} qualitatively characterizes the inference: constraints are improved either with more observations or with better measurements.
The $M^{-2p}$ term further indicates that upper limits from heavier systems are less informative than numerically similar upper limits from lighter systems for $p>0$. 
The funnel-like structure in $\mu_0-p$ and $\sigma_0-p$ is driven by Eq.~\eqref{eq:prec} and leads the marginal for $p$ to prefer higher values (since lower values are disallowed by the data).
This feature is, however, prior-dominated and will remain so until
a deviation is detected and $(\mu_0,\sigma_0)=(0,0)$ is excluded.

In Fig.~\ref{fig:ppds} we plot the distribution of deviations that are consistent with observations for each PN order, i.e., the posterior predictive distribution,
\begin{equation}
    p(\delta\varphi_k | M, d) = \int\textrm{d}\Lambda\, p(\Lambda|d)\,  \pi(\delta\varphi_k|\Lambda,M)\,,
    \label{eq:PPD}
\end{equation}
where $\Lambda \equiv \{\mu_0,\sigma_0,p\}$, $\pi(\delta\varphi_k|\Lambda, M)$ is the deviation Gaussian distribution, $p(\Lambda|d)$ is the posterior on $\Lambda$ (cf., Fig.~\ref{fig:gwtc3_corner}), and $d$ is the data. 
The integral is computed by averaging Gaussian distributions $\pi(\delta\varphi_k|\Lambda, M)$ over the posterior $p(\Lambda|d)$. We present Eq.~\eqref{eq:PPD} for $p=0, 4, 6$ (blue, purple, and orange) as well as integrating over $p$ (shaded). 
The distributions areconsistent with GR, \change{$\mu_0=\sigma_0=0$, constrained within the ${\cal{Q}}_\textrm{GR}<46\%$ credible intervals for all PN orders}.
We find overall similar behavior across orders.
In each panel, the upper sub-panel shows the precision \change{${\cal{P}}(M,p)$}, \change{normalized independently for each index $p$. For all cases of $p$, the precision is maximized} at ${\sim}20\,M_\odot$.
For $p\geq4$, corresponding to corrections for gravity in 4-dimensional spacetimes~\cite{Sotiriou_2014,Endlich:2017tqa,deRham:2021bll}, constraints are dominated by lower total mass binaries.

\section{Detectability of simulated violations}

\begin{figure}
    \centering
    \includegraphics[width=\linewidth]{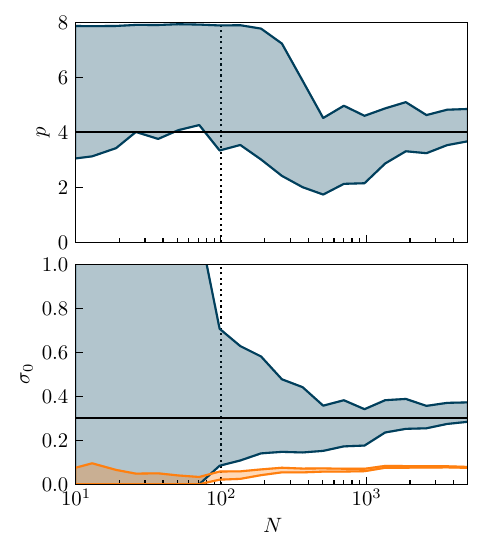}
    \caption{Inferred curvature scaling $p$ (top) and standard deviation $\sigma_0$ (bottom) at the 90\% level as a function of the number of simulated GW observations.
    The blue bounds correspond to an analysis that infers the curvature index, $p$, whereas the orange corresponds to fixing $p=0$.
    The true values are shown in \change{solid} black \change{horizontal} lines.
    For this population we infer a violation of GR, i.e., $\sigma_0>0$, starting at $N\sim100$ (\change{dotted black} vertical line), while $p=0$ and $6$ are ruled out by the data after $N\sim 500$ observations.
    Fixing $p=0$ misestimates the deviation. }
    \label{fig:injections}
\end{figure}

The current GW catalog does not exhibit evidence of a deviation from GR, we therefore explore inference in the presence of deviations with a simulated catalog of $N=5000$ observations.
We consider the 0PN order and simulate data with $\mu_0=0$, $\sigma_0=0.3$, $p=4$ per Eq.~\eqref{eq:musigma}, which is consistent with current constraints.
For simplicity, we adopt a mass distribution that matches the current observations and apply no selection effects.
With this simulated catalog, we repeat the analysis and present 90\% constraints on $\sigma_0$ and $p$ for varying numbers of detections in Fig.~\ref{fig:injections} (blue).
For reference, we compare to an analysis that fixes $p=0$ (orange), corresponding to the standard procedure of Refs.~\cite{Payne:2023kwj,LIGOScientific:2021sio}.

Fewer observations are required to identify a deviation from GR ($\sigma_0 >0$) than to
constrain its curvature scaling. For these simulations,
$\sigma_0 = 0$ is excluded at the $90\%$ level
after ${\sim}100$ observations, whereas data-driven (as opposed to prior-dominated---c.f., discussion of Fig.~\ref{fig:gwtc3_corner}) constraints on $p$ require
${\cal{O}}(500)$ observations (blue). A model without curvature scaling (fixing $p=0$) identifies a violation of GR with a similar number of observations
but provides no information about its curvature scaling and infers a
lower value of $\sigma_0\sim0.08$ (orange). The addition of the curvature
dependence in the inference unlocks the capability to infer the curvature
structure and characterize the properties of a putative deviation. In
this example, we would be able to rule out propagation effects ($p=0$) and
quartic curvature corrections ($p=6$) after ${\sim}500$ observations.
Although these exact numbers depend on the mass distribution and simulated deviation, we expect the general trends to be robust.

\section{Conclusions}

In this Letter, we have extended tests of GR with GW inspirals to incorporate physical expectations for the curvature dependence of extensions of GR.
This approach not only incorporates more physically realistic---albeit still theory-agnostic---models, but also allows us to better characterize the nature of the deviation by inferring its scaling with spacetime curvature.
We applied this method to existing LIGO-Virgo-KAGRA observations, finding consistency with GR.
We also demonstrated, with simulated signals, how the curvature dependence can be constrained and thus provide clues about the properties of the beyond-GR theory.
Although we focused on PN inspiral deviations, this method can be applied to any test with a dimensionless deviation parameter.
\change{More broadly, the key realization of this work, namely that the curvature scaling can be agnostically inferred from data, can be leveraged across all tests of GR --- beyond the field of GW astronomy.}
\change{Beyond GR, this agnostic approach can be tailored to any effective-field-theory treatment, e.g., to analyze the temperature-dependent scaling of transport coefficients affecting viscous effects in hydrodynamics~\cite{Bemfica_2019}.}

Beyond constraining the curvature dependence, our physical arguments
suggest ways to either strengthen confidence in a detected deviation or
safeguard against systematics, e.g.,~\cite{Gupta:2024gun}. 
Firstly, if GR is found to be incorrect and the curvature scaling $p$ is inferred to be an integer, it will immediately inform on viable theories.
Further, extraction of $p$ at different PN orders would not only allow for
consistency checks, but also---in case of differences---to draw key information on potential theories.
For example, it is possible that some specific PN corrections are subleading, displaying a higher curvature scaling than the majority of the other PN corrections due to the underlying
model under consideration (e.g., no dipole radiation for equal-mass
objects in scalar-tensor theories). 
In all cases, different tests (e.g., PN phase and ringdown) should give compatible results.
This idea can further be extended to multiparameter tests~\cite{Gupta:2020lxa, Datta:2022izc}. 

Secondly, false deviations could be induced by missing
physics~\cite{Gupta:2024gun}, waveform systematics~\cite{Purrer:2019jcp,
Moore:2021eok}, or detector glitches~\cite{Kwok:2021zny}. These effects often
have specific mass-dependent behavior, e.g., $M^{-5/6}$ for
eccentricity~\cite{Saini:2022igm, Saini:2023rto} or large deviations only present for
heavy binary masses due to glitches~\cite{Ashton:2021tvz}.
Therefore constraining the
mass dependence would help distinguish between such systematics and a genuine GR deviation under the effective-field-theory framework.

Finally, the expectation that \change{$p\geq4$} suggests that the highest-curvature black holes,
i.e., the \emph{lightest} black holes, will yield the strongest constraints{\footnote{\change{For non-vacuum systems, the expected value is $p=2$ as quadratic curvature corrections 
would be the leading modification in the EFT action. However, such events represent a more difficult challenge for testing GR with lower observational rates and the introduction of matter effects.}}. For
example, a ${\cal O}(0.1)$ deviation constraint from a $10\,M_\odot$ binary is
equivalent to a ${\cal O}(10^{-21})$ constraint from a $10^6\,M_\odot$ system if
$p=4$---the expectation for cubic or quadratic corrections with an additional
degree of freedom. This suggests that ground-based GW detectors, including the
next-generation Einstein Telescope~\cite{Maggiore:2019uih, Branchesi:2023mws}
and Cosmic Explorer~\cite{Evans:2021gyd, Evans:2023euw} detectors, will provide
deeper probes of GR than observations of supermassive black holes with pulsar
timing arrays~\cite{NANOGrav:2021ini,NANOGrav:2023ygs}, the Event Horizon
Telescope~\cite{2009astro2010S..68D, 2009ApJ...695...59D} or LISA (beyond the
extreme mass ratio regime)~\cite{LISA, Gair:2012nm}. Modeling the curvature
dependence within these GW tests allows us to more deeply probe the fundamental
nature of gravity and/or invalidate whole families of theories without resorting to theory-specific models.

\section{Acknowledgements}

We thank Cliff Burgess, Guillaume Dideron, Jaume Gomis, Isaac Legred, 
 Leah Jenks and Rafael Porto for discussions.
EP was supported by NSF Grant PHY-2309200.
KC was supported by NSF Grant PHY-2110111.
LL acknowledges support from the Natural Sciences and Engineering Research Council of Canada through a Discovery Grant. LL also thanks financial support via
the Carlo Fidani Rainer Weiss Chair at Perimeter Institute and CIFAR.
This research was supported in part by Perimeter Institute for Theoretical Physics. Research at Perimeter Institute is supported in part by the Government of Canada through the Department of Innovation, Science and Economic Development and by the Province of Ontario through the Ministry of Colleges and Universities.
YC was supported by the Simons Foundation (Award Number 568762) and NSF Grants PHY-2011961 and PHY-2011968. 
Computing resources were provided by the Flatiron Institute. The Flatiron Institute is funded by the Simons Foundation.

This material is based upon work supported by NSF's LIGO Laboratory which is a major facility fully funded by the National Science Foundation.
This research has made use of data, software and/or web tools obtained from the Gravitational Wave Open Science Center (https://www.gw-openscience.org), a service of LIGO Laboratory, the LIGO Scientific Collaboration and the Virgo Collaboration.
Virgo is funded by the French Centre National de Recherche Scientifique (CNRS), the Italian Istituto Nazionale della Fisica Nucleare (INFN) and the Dutch Nikhef, with contributions by Polish and Hungarian institutes.
The authors are grateful for computational resources provided by the LIGO Laboratory and supported by National Science Foundation Grants PHY-0757058 and PHY-0823459.
This manuscript carries LIGO Document Number \#P2400277.

\change{We built this analysis upon Refs.~\cite{Payne:2023kwj, code_release}, utilizing {\sc NumPyro}~\citep{phan2019composable, bingham2019pyro} and {\sc JAX}~\citep{jax2018github}. 
We use {\tt \sc AstroPy}~\cite{astropy:2013, astropy:2018, astropy:2022} and {\tt \sc SciPy}~\cite{Hunter:2007} for additional calculations, and plot the results with {\tt \sc corner}~\cite{corner} and {\tt \sc arViz}~\cite{arviz_2019}.} 

\bibliography{refs}

\end{document}